\documentclass[a4paper,10pt,twoside]{cpc-hepnp}
\usepackage{multicol}
\usepackage{graphicx}
\usepackage{booktabs}
\usepackage{amssymb,bm,mathrsfs,amscd}
\usepackage[tbtags]{amsmath}
\usepackage{lastpage}
\usepackage{braket,slashed}
\usepackage{multirow}
\usepackage{CJK}
\newcommand{\MeV}{\text{MeV}}
\newcommand{\GeV}{\text{GeV}}

\newcommand{\dif}{\text{d}}
\newcommand{\me}{\text{e}}

\begin{document}
\begin{CJK*}{GBK}{song}

\fancyhead[co]{\footnotesize Fine Splittings in Charmonium Spectrum
  with Channel Coupling Effect} 
\footnotetext[0]{Received \today}

\title{Fine splitting in charmonium spectrum with channel coupling
  effect\thanks{Supported by the National Natural Science Foundation
    of China under Grant 10675008.}}
  
\author{YANG Chun$^1$%
  \quad LI Bao-Fei$^1$%
  \quad CHEN Xiao-Lin$^2$%
  \quad DENG Wei-Zhen$^{1;1)}$\email{dwz@pku.edu.cn}}
\maketitle

\address{%
  1~(Department of Physics and State Key
  Laboratory of Nuclear Physics and Technology, \\
  Peking University, Beijing 100871, China)\\
  2~(Department of Physics, Peking University, Beijing 100871, China)\\
}

\begin{abstract}
  We study the fine splitting in charmonium spectrum in the quark model
  with the channel coupling effect, including $DD$, $DD^*$, $D^*D^*$
  and $D_sD_s$, $D_sD_s^*$, $D_s^*D_s^*$ channels. The interaction for
  channel coupling is constructed from the current-current Lagrangian
  related to the color confinement and the one-gluon exchange
  potentials.  By adopting the massive gluon propagator from the
  lattice calculation in the nonperturbative region, the coupling
  interaction is further simplified to the four-fermion interaction.
  The numerical calculation still prefers the assignment $1^{++}$ of
  $X(3872)$.
\end{abstract}

\begin{keyword}
  quark model, four-fermion interaction, coupled-channel, $X(3872)$
\end{keyword}

\begin{pacs}
  12.39.Jh, 12.39.Pn, 14.40.Lb
\end{pacs}
\end{CJK*}

\begin{multicols}{2}

\section{Introduction}

A series of hidden charm states, the so-called $X$, $Y$, $Z$, have been
discovered and confirmed by the experiments since 2003. The nature of
these narrow resonances has attracted much attention, because their
properties are not consistent with the prediction of the quark model.

The typical $X(3872)$ state, which was discovered in 2003 by the Belle
Collaboration \cite{Choi:2003ue} and subsequently confirmed by the CDF
Collaboration \cite{Acosta:2003zx} and BABAR Collaboration
\cite{Aubert:2004ns}, etc., is now listed with $M_X=3872.2\pm0.8\MeV$,
$\Gamma_X=3.0^{+1.9}_{-1.4}\pm0.9\MeV$ in
PDG\cite{Amsler:2008zzb}. Its quantum numbers were inferred
$J^{PC}=1^{++}$ or $2^{-+}$. The corresponding charmonium candidate in
the quark model is $2^3P_1$ or $1^1D_2$ respectively.

The mass of the $2^3P_1$ state in the quark model is $\sim 100$ MeV
above $M_X$. However, the channel coupling effects by the creation of
open charmed meson pairs can produce significant mass shift to the
bare charmonium spectrum.  In Ref.~\cite{Eichten:2004uh}, only the
fine splitting in the mass shift induced by open-charm states is
considered. In Refs.~\cite{Barnes:2004fs,Kalashnikova:2005ui}, the
whole mass shift is considered to lower the bare mass of the excited
charmonium state. The mass shift can be also handily treated by
introducing screened potential into the quark model\cite{Li:2009ad}.

The proximity of the $X(3872)$ to $DD^*$ threshold implies that the
cusp scenario may be important \cite{Bugg:2004rk}. The cusp can be
calculated from channel coupling and the result is in qualitative
agreement with experiment \cite{Danilkin:2010cc}.  The observed but
Okubo-Zweig-Iizuka (OZI) forbidden decay channel $\rho J/\psi$ is also
considered in Ref.~\cite{Coito:2010if}.

Recently, a study of the $\pi^+\pi^-\pi^0$ mass distribution from the
$X(3872)$ decay by the BABAR Collaboration favors the negative
parity assignment $2^{-+}$ \cite{delAmoSanchez:2010jr}.
However, the mass of the corresponding charmonium state $1^1D_2$ in
the quark model is $\sim 100$ MeV below $M_X$. Since the $\psi(3770)$
is assigned to $1^3D_1$ in the quark model, the assignment $2^{-+}$ seems
to conflict with the small fine splitting in $c\bar{c}$ $1D$ multiplet
from the quark model calculation \cite{Barnes:2003vb}.

The mechanism of channel coupling is the same as strong decay's.  The
simplest decay model is the so-called ${}^3P_0$ model based on the
flux-tube-breaking model \cite{Micu:1968mk, LeYaouanc:1972ae}.
Another model is the Cornel model which tries to relate the pair-creation
interaction to the potential in the quark model \cite{Eichten:1978tg,
  Eichten:1979ms}.  The Cornel model assumes the Lorentz vector
confinement so the total vector potential is
\begin{equation}
  V(r)=-\frac{\kappa}{r} + \frac{r}{a^2}.
\end{equation}
Thus in the Cornel model the decay amplitude from the one-gluon exchange
and that from the confinement add destructively. A similar calculation
but using the Lorentz scalar confinement shows that the decay
amplitude from the scalar linear confinement is too large
\cite{Ackleh:1996yt}.

The lattice calculation shows that the gluon propagator is quite
different in the nonperturbative region. The gluon may get a mass
about $600\sim 1000$ MeV
\cite{Leinweber:1998uu,Silva:2004bv,Oliveira:2009nn}.
A non-vanishing gluon mass is used in the phenomelogical calculation
of the diffractive scattering\cite{Forshaw:1998tf} and
radiative decays of the $J/\psi$ and $\Upsilon$\cite{Field:2001iu}.

In this work, we will consider the fine splitting induced by channel
coupling with open-charm states, including $DD$, $DD^*$, $D^*D^*$ and
$D_sD_s$, $D_sD_s^*$, $D_s^*D_s^*$.  Following the Cornel model, we will
construct the model pair-creation interaction from the potential in
the quark model, i.e. the scalar confinement plus the vector
one-gluon exchange. With the assumption of the massive gluon
propagator in the pair-creation process, we will obtain a simple
effective four-fermion interaction which is quite similar to the case
of weak interaction. In Sec.~2, we will introduce the channel coupling
model. In Sec. 3, the numerical analysis is performed. Finally, we
will give a brief summary.

\section{The channel coupling model}

In the simplest version of channel coupling model
\cite{Kalashnikova:2005ui}, the hadronic state is assumed to be
represented by
\begin{equation}
  \ket{\Psi_\alpha} = \begin{pmatrix}
    c_\alpha \ket{\psi_\alpha} \\
    \sum_i \chi_{\alpha i} \ket{M_1(i)M_2(i)}
  \end{pmatrix},
\end{equation}
where the bare state $\ket{\psi_\alpha}$ is coupled to several
meson-meson channels $\ket{M_1(i)M_2(i)}$. The system Hamiltonian reads
\begin{equation}
  \hat{H} = \begin{pmatrix}
    \hat{H}_c & \hat{V} \\
    \hat{V} & \hat{H}_{M_1M_2}
  \end{pmatrix},
\end{equation}
where $\hat{H}_c$ is the meson Hamiltonian of the quark model, with
\begin{equation}
  \hat{H}_c \ket{\psi_\alpha}=M_\alpha \ket{\psi_\alpha}.
\end{equation}
In this work, $\hat{H}_{M_1M_2}$ includes only the free meson
Hamiltonian, so
\begin{equation}
  \hat{H}_{M_1M_2} = \hat{H}_{M_1}+\hat{H}_{M_2} .
\end{equation}

The Hamiltonian in the non-relativistic quark potential model can always
be written as
\cite{Kalashnikova:2005ui}
\begin{equation}
  \hat{H}_c=\hat{H}_0 + \hat{H}_{sd},
\end{equation}
where $\hat{H}_0$ and $\hat{H}_{sd}$ are the spin-independent and
spin-dependent parts respectively. The spin-independent part reads
\begin{equation}
  \hat{H}_0 = \frac{p^2}{2\mu} + V(r) + C,
\end{equation}
$\mu$ is the reduced mass. The potential $V(r)$ is usually taken to be
a sum of the linear confinement plus the one-gluon exchange Coulomb
potential:
\begin{equation}
  V(r)=\sigma r - \frac43 \frac{\alpha_s}{r}.
\end{equation}
$\hat{H}_{sd}$ includes spin-spin, spin-orbit and tensor force:
\begin{equation}
  H_{sd}=V_{HF}(r)\bm{S}_1\cdot\bm{S}_2
  +V_{LS}(r)\bm{L}\cdot\bm{S}
  +V_T(r) T ,
\end{equation}
which determines the fine splitting in the spectrum.

The off-diagonal interaction $\hat{V}$ is responsible for channel
coupling. It depends on the pair-creation mechanism of the specific
hadron decay model.  The ${}^3P_0$ model \cite{Micu:1968mk,
  LeYaouanc:1972ae} and the Cornel model \cite{Eichten:1978tg,
  Eichten:1979ms} are two popular decay models.  

To describe the creation of a light-quark pair in the quark model, a
plausible approach is to consider the quantum field expression of the
quark potential $V(r)$.  In the Cornell model, the quark potential is
replaced by an instantaneous interaction \cite{Eichten:1978tg,
  Eichten:1979ms}
\begin{equation}
  H_I = \frac12 \int \dif^3 x \dif^3 y :\rho_a(\bm{x}) \frac34
  V(\bm{x}-\bm{y}) \rho_a(\bm{y}): ,
\end{equation}
where
\begin{equation}
  \rho_a(\bm{x}) = \sum_{\text{flavors}} \psi^\dag(\bm{x}) \frac12
  \lambda_a \psi(\bm{x}),
\end{equation}
is the quark color-charge-density operator, and $\psi(\bm{x})$ is the
quark field operator. As the spin splitting in charmonium spectrum and
the lattice gauge calculation indicate that the confinement current
should be the Lorentz scalar, in Ref.~\cite{Ackleh:1996yt} the
instantaneous interaction is replaced by the scalar confinement
interaction plus the vector one-gluon exchange.

Following the Cornel model, here we will model the pair-creation from the
quark model.  We first assume the nonlocal current-current action of
the quark interaction \cite{Godfrey:1985xj}:
\end{multicols}
\ruleup
\begin{align}
  A =& -\frac12\int \dif^4 x \dif^4 y \bar{\psi}(x) \gamma_\mu \frac12
  \lambda_a\psi(x)  G(x-y) \bar{\psi}(y) \gamma^\mu\frac12
  \lambda_a\psi(y) \notag\\
  & -\frac12\int \dif^4 x \dif^4 y \bar{\psi}(x) \frac12
  \lambda_a\psi(x) S(x-y) \bar{\psi}(y) \frac12 \lambda_a\psi(y) .
\end{align}
\ruledown \vspace{0.5cm}
\begin{multicols}{2}
  The vector kernel $G$ is obtained from the one-gluon propagator. In
  the momentum space
\begin{equation}
  G(q^2) = -\frac{4\pi\alpha_s}{q^2} .
\end{equation}
The scalar kernel $S(x-y)$ is obtained from the linear confinement
\begin{equation}
  S(q^2) = -\frac{6\pi b}{q^4} .
\end{equation}

The lattice calculation shows that the behavior of the gluon
propagator is quite different in the nonperturbative region. The gluon
may get a mass about $600\sim 1000$ MeV
\cite{Leinweber:1998uu,Silva:2004bv,Oliveira:2009nn}.  With the gluon
getting a mass in the nonperturbative region, we can make the
non-relativistic approximation $q^2 \to q^2-m_g^2\approx -m_g^2$ in
the quark-antiquark pair-creation process. Thus
\begin{align}
  D_{\mu\nu}(q^2) \approx& \frac{4\pi\alpha_s g_{\mu\nu}}{m_g^2}, \\
  D(q^2) \approx& -\frac{6\pi b}{m_g^4}.
\end{align}
Then the channel coupling interaction is simplified to the
four-fermion interaction
\begin{align}
  \hat{V} =& -\frac12 \frac{4\pi\alpha_s }{m_g^2} 
  \int \dif^3 x \bar{\psi}(\bm{x}) \gamma_\mu \frac12
  \lambda_a\psi(\bm{x})  
  \bar{\psi}(\bm{x}) \gamma^\mu\frac12
  \lambda_a\psi(\bm{x}) \notag\\
  & +\frac12 \frac{6\pi b}{m_g^4} 
  \int \dif^3 x \bar{\psi}(\bm{x}) \frac12
  \lambda_a\psi(\bm{x}) 
  \bar{\psi}(\bm{x}) \frac12 \lambda_a\psi(\bm{x})
\end{align}


Once we calculate the transition amplitudes
\begin{equation}
  f_i(\bm{p}) = \braket{\psi_\alpha|\hat{V}|M_1(i)M_2(i)},
\end{equation}
where $\bm{p}$ is the relative momentum between $M_1$ and $M_2$,
the mass shifts are given by
\begin{align}
  g(M)=& \sum_i g_i(M), \\
  g_i(M)=& \int \frac{f_i(\bm{p}) f_i(\bm{p})}
  {\left(m_{i1}+m_{i2}+\frac{p^2}{2\mu_i} \right) - M} \dif^3 p,
\end{align}
where $m_{i1}$ and $m_{i2}$ are the masses of $M_1(i)$ and $M_2(i)$ mesons,
$\mu_i$ is their reduced mass.

To calculate the coupling matrix element, we will use the simple
harmonics oscillator (SHO) wave functions as usual.  The partial-wave
amplitude $f^{ls}$ can be expressed as
\begin{equation}
  \label{part-amp}
  f^{ls}(A\to BC)=\pi^{-\frac{7}{4}} \beta_A^{3/2} 
  \me^{-\frac{m^2_c}{2(m_q+m_c)^2(\beta^2_A+\beta^2_B)} p^2} F^{ls}(p),
\end{equation}
where $\beta_B=\beta_C$, $m_c$ is the mass of charm quark, $m_q$
is the mass of light quarks ($u$, $d$, or $s$).  $F^{ls}(p)$ is a
polynomial of $p$ which depends on the specific channel (the formulas
are collected in Appendix).

Our calculation is basically non-relativistic. But the exponential
factor in the obtained partial-wave amplitude Eq.~(\ref{part-amp}) is
obviously not enough to cut of the high momentum contribution. We will
make an additional cutoff to the momentum integration. The mass shift
is then replaced by
\begin{equation}
  \label{mass-shift}
  g_i(M)= \int \frac{f_i(\bm{p}) f_i(\bm{p})}
  {\left(m_{i1}+m_{i2}+\frac{p^2}{2\mu_i} \right) - M} 
  \exp(-p^2/\Lambda^2)\dif^3 p,
\end{equation}
where $\Lambda$ is the cutoff parameter.

Since the channel coupling calculation is essentially the virtual
charmed meson loop calculation, the quark potential in the quark model
should be renormalized \cite{Li:2009ad}. The renormalization process
can be outlined as follows. The full Hamiltonian is divided into
\begin{equation}
  \hat{H}_{full} = \hat{H}_c + \Delta \hat{H}.
\end{equation}
$\hat{H}_c$ is the original quark model Hamiltonian. Its spectrum
is given by
\begin{equation}
  \label{spectrum-formula}
  M_{nslj} = M_{nl} + \braket{V_{HF}} \braket{\bm{S}_1\cdot\bm{S}_2}
  + \braket{V_{LS}} \braket{\bm{L}\cdot\bm{S}}
  + \braket{V_T} \braket{T},
\end{equation}
where $M_{nl}$ is the centroid of $nl$ multiplet which is obtained
from the spin-independent Hamiltonian $\hat{H}_0$ and the remaining terms
give the fine splitting. $\braket{T}$ is the expectation value
of the tensor operator,
\begin{equation}
  \braket{T} = \begin{cases}
    -\frac16 \frac{l+1}{2l-1} & j=l-1, \\
    \frac16 & j=l, \\
    -\frac16 \frac{l}{2l+3} & j=l+1 ,
  \end{cases}
\end{equation}
where the total spin $s=1$.  $\Delta \hat{H}$ is the cancellation term
whose contribution should be added to the mass shift from
coupled-channels to give the renormalized mass shift. The renormalized
mass shift contains both a centroid correction and a fine splitting
one. The centroid contribution will modify the quark central potential
\cite{Li:2009ad}. It is the fine splitting correction we will consider
in this work.

\section{Numerical calculation of fine splitting}

In our calculation, the quark model is taken from
Ref.~\cite{Kalashnikova:2005ui}.  The potential parameters are:
\begin{align}
  &\alpha_s=0.55, &&\sigma = 0.175\GeV^2, &&m_c=1.7\GeV, \notag\\
  &C=-0.271\GeV, && m_q=0.33\GeV, &&m_s=0.5\GeV.
\end{align}
The SHO parameter $\beta$ is determined from the mean square radius of
the meson state.  The $\beta$ values of open-charm states are
\begin{align}
  &\beta_D=0.385\GeV, &&\beta_{D_S}=0.448\GeV,
\end{align}
and the $\beta$ values of charmonium states are listed in
Table~\ref{table1}.
\begin{center}
\tabcaption{\label{table1}%
  The $\beta$ values of charmonium states.}
\begin{tabular*}{80mm}{cccccc}
  \hline\hline
  $nL$ & $1S$ & $2S$ & $1P$ & $2P$ & $1D$ \\\hline
  $\beta(\GeV)$ & $0.676$ & $0.485$ & $0.514$ 
  & $0.435$ & $0.461$ \\\hline\hline
\end{tabular*}
\end{center}
In our calculation we take the gluon mass $m_g=640\MeV$. This gives
\begin{equation}
  \Gamma(\psi(3770)\to D\bar{D})=28.2 \MeV,
\end{equation}
to fit the expermental value $27.3\pm1.0 \MeV$ \cite{Amsler:2008zzb}.

To calculate the mass shift, we need further to know the physical mass
$M$ in Eq.~(\ref{mass-shift}).  For the charmonium $1S$, $1P$ and $2S$
multiplets, we can directly use the experimental masses from PDG
\cite{Amsler:2008zzb}. For the $2P$ and $1D$ multiplets, the physical
masses are the predicted values calculated from the assignments of
$\psi(3770)$ to $1^3D_1$ and $X(3872)$ to $2^3P_1$.

The mass shifts are listed in Table~\ref{table2}.  In our calculation
we take the cutoff paramter $\Lambda = 800\MeV$. We also show the mass
shifts without the integration cutoff. The cutoff reduces the mass
shift by $\sim 15\%$, which means that the contribution from high
transfer momentum will be about $85\%$ if we do not make the cutoff in
this non-relativistic calculation.
\end{multicols}
\begin{center}
  \tabcaption{\label{table2}%
    The mass shifts of charmonium states in MeV. The last column lists
    the total mass shifts without the integration cutoff.}
\begin{tabular*}{170mm}{@{\extracolsep{\fill}}|c|r|r|r|r|r|r|r|r|}
\hline\hline
$n^{2S+1}L_J$ & $DD$ & $DD^*$ & $D^*D^*$ & $D_sD_s$ & $D_sD^*_s$
& $D^*_s D^*_s$ & total & no cutoff \\\hline
$1^3S_1$ & $-9$ & $-36$ & $-64$ & $-6$ & $-26$ & $-49$ & 
$-190$ & $-1359$ \\
$1^1S_0$ & $0$ & $-52$ & $-47$ & $0$ & $-39$ & $-36$ & 
$-175$ & $-1274$ \\
$1^3P_2$ & $-12$ & $-32$ & $-75$ & $-5$ & $-15$ & $-37$ & 
$-175$ & $-1035$ \\
$1^3P_1$ & $0$ & $-53$ & $-52$ & $0$ & $-21$ & $-26$ & 
$-152$ & $-1021$ \\
$1^3P_0$ & $-23$ & $0$ & $-67$ & $-7$ & $0$ & $-34$ & 
$-131$ & $-968$ \\
$1^1P_1$ & $0$ & $-61$ & $-50$ & $0$ & $-27$ & $-24$ & 
$-162$ & $-1021$ \\
$2^3S_1$ & $-6$ & $-18$ & $-31$ & $-1$ & $-4$ & $-8$ & 
$-68$ & $-872$ \\
$2^1S_0$ & $0$ & $-28$ & $-21$ & $0$ & $-7$ & $-6$ & 
$-62$ & $-839$ \\
$2^3P_2$ & $-1$ & $-9$ & $-16$ & $-1$ & $-3$ & $-7$ & 
$-37$ & $-691$\\
$2^3P_1$ & $0$ & $-17$ & $-10$ & $0$ & $-4$ & $-4$ & 
$-35$ & $-716$ \\
$2^3P_0$ & $-5$ & $0$ & $-13$ & $-1$ & $0$ & $-5$ & 
$-25$ & $-680$ \\
$2^1P_1$ & $0$ & $-18$ & $-10$ & $0$ & $-5$ & $-4$ & 
$-36$ & $-701$ \\
$1^3D_3$ & $-8$ & $-18$ & $-49$ & $-2$ & $-5$ & $-15$ & 
$-98$ & $-652$ \\
$1^3D_2$ & $0$ & $-40$ & $-33$ & $0$ & $-9$ & $-11$ & 
$-93$ & $-665$ \\
$1^3D_1$ & $-28$ & $-14$ & $-38$ & $-2$ & $-3$ & $-13$ & 
$-98$ & $-669$ \\
$1^1D_2$ & $0$ & $-44$ & $-31$ & $0$ & $-11$ & $-9$ & 
$-95$ & $-657$\\\hline\hline
\end{tabular*}
\end{center}
\begin{multicols}{2}
  The fine splittings are listed in Table~\ref{table3}. For $1S$,
  $1P$, $2S$ states, the physical mass is the experimental mass. Then
  the fine splitting is calculated for each multiplet and listed as
  ``splitting required''. The fine splitting from the quark model is
  calculated from the bare masses of the quark model which is also taken
  from Ref.~\cite{Kalashnikova:2005ui}. The fine splitting from
  coupled-channels are listed in the last column. So the total model
  fine splitting is the sum of the contributions from the quark model
  and from the coupled-channels. The results show that the calculated
  splittings fit the ``splitting required'' well in $1S$ and $2S$
  multiplets. However in the $1P$ multiplet, the model splittings seem
  too large.

\begin{center}
  \tabcaption{\label{table3}%
    The physical masses and fine splittings.}
  \begin{tabular*}{80mm}{@{\extracolsep{\fill}}|c|c|r|r|r|}
\hline\hline
\multirow{2}{*}{$n^{2S+1}L_J$} & \multirow{2}{*}{mass} & 
\multicolumn{1}{c|}{splitting} & \multicolumn{1}{c|}{splitting} & 
\multicolumn{1}{c|}{splitting} \\
&& \multicolumn{1}{c|}{required} & \multicolumn{1}{c|}{q. m.} & 
\multicolumn{1}{c|}{c. c.} \\\hline
$1^3S_1$ & $3097$ & $+29$ & $+32$ & $-4$ \\
$1^1S_0$ & $2980$ & $-87$ & $-97$ & $+12$ \\\hline
$1^3P_2$ & $3556$ & $+31$ & $+36$ & $-13$ \\
$1^3P_1$ & $3511$ & $-15$ & $-19$ & $+11$ \\
$1^3P_0$ & $3415$ & $-110$ & $-106$ & $+31$ \\
$1^1P_1$ & $3525$ & $+0$ & $-5$ & $+0$ \\\hline
$2^3S_1$ & $3686$ & $+12$ & $+14$ & $-2$ \\
$2^1S_0$ & $3637$ & $-37$ & $-41$ & $+5$ \\\hline
$2^3P_2$ & $3918$ & $+30$ & $+32$ & $-2$ \\
$2^3P_1$ & $3872$ & $-17$ & $-17$ & $+0$ \\
$2^3P_0$ & $3808$ & $-80$ & $-90$ & $+10$ \\
$2^1P_1$ & $3881$ & $-7$ & $-6$ & $-1$ \\\hline
$1^3D_3$ & $3798$ & $+6$ & $+8$ & $-2$ \\
$1^3D_2$ & $3795$ & $+3$ & $-0$ & $+3$ \\
$1^3D_1$ & $3773$ & $-19$ & $-17$ & $-2$ \\
$1^1D_2$ & $3793$ & $+0$ & $-0$ & $+1$ \\\hline\hline
\end{tabular*}
\end{center}

  Next, we turn to the $2P$ and $1D$ multiplets. This time, the
  ``required spltting'' is the sum of the splitting from the quark
  model and from the coupled-channels. For the $1D$ multiplet, the
  $\psi(3770)$ is assigned to the $1^3D_1$ state. Then the masses of
  other states in the multiplet are calculated from the fine
  splittings as the prediction. The predicted mass of $1^1D_2$ is
  $3793\MeV$. So the $c\bar{c}$ $1^1D_2$ state is unlikely to be the
  experimental $X(3872)$ state even when we have considered the fine
  splitting from coupled-channels.  So we assign the $X(3872)$ to the
  $2^3P_1$ state and calculate the masses of the rest states in the $2P$
  multiplet.
 
\section{Summary}

We have calculated the fine splitting in charmonium spectrum in the quark
model with the channel coupling effect. The open charmed meson-meson
channels below $4\GeV$, including $DD$, $DD^*$, $D^*D^*$ and $D_sD_s$,
$D_sD_s^*$, $D_s^*D_s^*$, are considered. The current-current nonlocal
interacting action is constructed from the color confinement and the
one-gluon exchange interaction in the quark model. Using the massive
gluon propagator from the lattice calculation in the nonperturbative
region, the coupling interaction is further simplified approximately
to the four-fermion interaction.  The numerical calculation still
prefers the assignment $1^{++}$ of $X(3872)$ after we consider the
fine splitting effect from the coupled-channels.  The $2P$ and $1D$
charmonium spectrums are estimated from the assignments of $1^3D_1$ to
$\psi(3770)$ and $2^2P_1$ to $X(3872)$.

\acknowledgments{We would like to thank professor Shi-Lin Zhu for useful
  discussions.}

\end{multicols}

\vspace{-1mm}
\centerline{\rule{80mm}{0.1pt}}
\vspace{2mm}

\begin{multicols}{2}

\def\url#1{}
\bibliographystyle{hepnp.bst}
\bibliography{refs}

\end{multicols}
\vspace{105mm}
\begin{multicols}{2}

\subsection*{Appendix}
\begin{small}
\noindent{\bf The Partial-Wave Amplitudes}

The partial-wave amplitude is the sum of contribution from the confinement
and from the coulomb interaction:
\begin{equation}
  F^{ls}=\frac{6\pi b}{m_g^4} F_\text{conf}^{ls} 
  - \frac{4\pi\alpha_s}{m_g^2} F_\text{coul}^{ls}.
\end{equation}
In the following, 
\begin{subequations}
\begin{align}
  D^{ij}_k =& \frac{\beta_A^i\beta_B^j}{(\beta_A^2+\beta_B^2)^{k/2}}, \\
  \xi_q = & \frac{m_q}{m_q+m_c}, \\
  \xi_c = & \frac{m_c}{m_q+m_c}.
\end{align}
\end{subequations}

For the confinement, $F^{ls}_\text{conf}$ can be represented as
\begin{equation}
  F^{ls}_\text{conf}=\frac1{m_q} F_l(p)C^{ls},
\end{equation}
where $C^{ls}$ is a spin-orbit recoupling coefficient
\begin{align}
  C^{ls}=&(-1)^{s_C+s+l_A+j_A}
  \begin{Bmatrix} s_A&s&1\\l&l_A&j_A
  \end{Bmatrix}
  \begin{Bmatrix} \frac12 &\frac12 & s_B\\
    \frac12& \frac12& s_C\\ s_A &1&s
  \end{Bmatrix} \notag\\
  &\times\sqrt{6(2s+1)(2l_A+1)(2s_A+1)(2s_B+1)(2s_C+1)}.
\end{align}
The $F_l(p)$ is the polynomial of transfer momentum $p$:
\end{small}
\end{multicols}
\ruleup
\begin{small}
\begin{align}
  F_p(1S\to1S+1S)=&-\frac{8}{3\sqrt{3}} (\xi_c D^{05}_5+2\xi_q D^{03}_3 )p \\
  F_p(2S\to1S+1S)=& \frac{4\sqrt{2}}{9} \left\{\left[\xi_c(7D^{25}_7-3D^{07}_7)
      +6\xi_q(D^{23}_5-D^{05}_5)\right] p - 2 \xi_c^2
    (\xi_cD^{25}_9 +2\xi_q D^{23}_7) p^3 \right\} \\
  F_s(1P\to1S+1S)=&-\frac{8\sqrt{2}}{9\sqrt{3}} \left[
    3 D^{15}_5 - \xi_c(\xi_cD^{15}_7 +2\xi_q D^{13}_5) p^2 \right] \\ 
  F_d(1P\to1S+1S)=&-\frac{16}{9\sqrt{3}}
  \xi_c(\xi_c D^{15}_7+2\xi_q D^{13}_5)p^2 \\
  F_s(2P\to1S+1S)=&\frac{8}{9\sqrt{15}} \left\{
    15(D^{35}_7 - D^{17}_7) 
    -5\xi_c\left[\xi_c(3D^{35}_9-D^{17}_9)+2\xi_q(D^{33}_7-D^{15}_7)\right]p^2 
    + 2\xi_c^3(\xi_c D^{35}_{11} + 2\xi_qD^{33}_9)p^4
    \right\} \\
  F_d(2P\to1S+1S)=&\frac{8\sqrt2}{9\sqrt{15}} \left\{
    \xi_c\left[\xi_c (9D^{35}_9-5D^{17}_9)+10\xi_q(D^{33}_7-D^{15}_7)\right]p^2
    -2\xi_c^3(\xi_cD^{35}_{11}+2\xi_q D^{33}_9) p^4 \right\} \\
  F_p(1D\to1S+1S)=&-\frac{16\sqrt{2}}{45}
  \left[5\xi_cD^{25}_7 p -\xi_c^2(\xi_c D^{25}_9+2\xi_q D^{23}_7)p^3 \right] \\
  F_f(1D\to1S+1S)=&-\frac{16}{15\sqrt3}
  \xi_c^2(\xi_c D^{25}_9 + 2\xi_q D^{23}_7) p^3
\end{align}
\ruledown
\vspace{0.5cm}
\begin{multicols}{2}
For the one-gluon exchange, $F^{ls}_{\text{coul}}$ is further decomposed to
\begin{equation}
  F^{ls}_{\text{coul}}=\frac1{m_q}F_{1l}(p)C^{ls}+\frac1{m_c}F_{2l}(p)C^{ls}
  -\frac1{m_c}F_{1l}(p)C_2^{ls},
\end{equation}
where $C_2^{ls}$ is another spin-orbit recoupling coefficient.
\begin{itemize} 
\item $s_A=s_B=s_C=1$
  \[
  C_2^{ls}=(-1)^{l_A+j_A} \sqrt{2(2s+1)(2l_A+1)}
  \begin{Bmatrix} l_A&1&l\\s&j_A&1
  \end{Bmatrix},
  \]
\item $s_A=1$, $s_B=s_C=0$
  \[
  C_2^{l=j_A,s=0}=-\sqrt{\frac{2(2l_A+1)}{2j_A+1}},
  \]
\item $s_A=s_B=1$, $s_C=0$
  \[
  C_2^{l,s=1}=(-1)^{l_A+j_A+1}\frac{\sqrt{3(2l_A+1)}}2
  \begin{Bmatrix} 1&1&1 \\ l&l_A&j_A
  \end{Bmatrix},
  \]
\item $s_A=0$
  \[
  C_2^{l,s=1}=0.
  \]
\end{itemize}
The polynomials $F_{1l}(p)$ and $F_{2l}(p)$ are:
\end{multicols}
\ruleup
\begin{align}
  \allowdisplaybreaks
  F_{1p}(1S\to1S+1S)=&\frac{8}{3\sqrt{3}} \xi_c 
  (D^{03}_3-D^{23}_5) p \\
  F_{2p}(1S\to1S+1S)=&\frac{8}{3\sqrt{3}} \xi_c 
  (D^{03}_3+D^{23}_5) p \\
  F_{1p}(2S\to1S+1S)=& -\frac{4\sqrt{2}}{9} \left[
    \xi_c(7D^{25}_7-3D^{43}_7
    +3D^{23}_5-3D^{05}_5) p + 2 \xi_c^3
    (D^{43}_9  - D^{23}_7) p^3 \right] \\
  F_{2p}(2S\to1S+1S)=& \frac{4\sqrt{2}}{9} \left[
    \xi_c(7D^{25}_7-3D^{43}_7
    -3D^{23}_5+3D^{05}_5) p + 2 \xi_c^3
    (D^{43}_9  + D^{23}_7) p^3 \right] \\
  F_{1s}(1P\to1S+1S)=&\frac{8\sqrt{2}}{9\sqrt{3}} \left[
    3 D^{15}_5 + \xi_c^2(D^{33}_7 - D^{13}_5) p^2 \right] \\ 
  F_{2s}(1P\to1S+1S)=&-\frac{8\sqrt{2}}{9\sqrt{3}} \left[
    3 D^{15}_5 + \xi_c^2(D^{33}_7 + D^{13}_5) p^2 \right] \\ 
  F_{1d}(1P\to1S+1S)=&-\frac{16}{9\sqrt{3}}
  \xi_c^2( D^{33}_7- D^{13}_5)p^2 \\
  F_{2d}(1P\to1S+1S)=&\frac{16}{9\sqrt{3}}
  \xi_c^2( D^{33}_7+ D^{13}_5)p^2 \\
  F_{1s}(2P\to1S+1S)=&-\frac{8}{9\sqrt{15}} \left[
    15(D^{35}_7 - D^{17}_7) 
    -5\xi_c^2(3D^{35}_9+D^{53}_9-D^{33}_7-D^{15}_7)p^2 
    - 2\xi_c^4(D^{53}_{11} - D^{33}_9)p^4
    \right] \\
  F_{2s}(2P\to1S+1S)=&\frac{8}{9\sqrt{15}} \left[
    15(D^{35}_7 - D^{17}_7) 
    -5\xi_c^2(3D^{35}_9-D^{53}_9-D^{33}_7+D^{15}_7)p^2 
    - 2\xi_c^4(D^{53}_{11} + D^{33}_9)p^4
    \right] \\
  F_{1d}(2P\to1S+1S)=&-\frac{8\sqrt2}{9\sqrt{15}} \left[
    \xi_c^2 (9D^{35}_9-5D^{15}_7+5D^{33}_7-5D^{53}_9) p^2
    +2\xi_c^4(D^{53}_{11}-D^{33}_9) p^4 \right] \\
  F_{2d}(2P\to1S+1S)=&\frac{8\sqrt2}{9\sqrt{15}} \left[
    \xi_c^2 (9D^{35}_9+5D^{15}_7-5D^{33}_7-5D^{53}_9) p^2
    +2\xi_c^4(D^{53}_{11}+D^{33}_9) p^4 \right] \\
  F_{1p}(1D\to1S+1S)=&\frac{16\sqrt{2}}{45}
  \left[5\xi_cD^{25}_7 p +\xi_c^3( D^{43}_9 - D^{23}_7)p^3 \right] \\
  F_{2p}(1D\to1S+1S)=&-\frac{16\sqrt{2}}{45}
  \left[5\xi_cD^{25}_7 p +\xi_c^3( D^{43}_9 + D^{23}_7)p^3 \right] \\
  F_{1f}(1D\to1S+1S)=&-\frac{16}{15\sqrt3}
  \xi_c^3( D^{43}_9 - D^{23}_7) p^3 \\
  F_{2f}(1D\to1S+1S)=&\frac{16}{15\sqrt3}
  \xi_c^3( D^{43}_9 + D^{23}_7) p^3
\end{align}
\end{small}
\end{document}